\documentstyle[preprint,aps,prabib]{revtex}
\begin{document}
\draft
\title{Quantization of spherically symmetric solution
of  SU(3) Yang-Mills theory}
\author{V. Dzhunushaliev
\thanks{E-mail address: dzhun@freenet.bishkek.su}}
\address{Theor. Physics Dept.,
Kyrgyz State National University,
720024, Bishkek, Kyrgyzstan}
\author{D. Singleton
\thanks{E-mail address : dasingle@maxwell.phys.csufresno.edu}}
\address{Dept. of Physics, CSU Fresno, 2345 East San Ramon Ave.,
M/S 37, Fresno, CA 93740-8031}
\date{\today}
\maketitle
\begin{abstract}
A recent investigation of the SU(3) Yang-Mills field equations 
found several classical solutions  which exhibited 
a type of confinement due to gauge fields which increased
without bound as $r \rightarrow \infty$. This increase 
of the gauge fields gave these solutions an infinite
field energy, raising questions about their physical
significance. In this paper we apply some ideas of Heisenberg
about the quantization of strongly interacting, non-linear fields
to this classical solution and find that at large $r$ this
quantization procedure softens the unphysical behaviour of the
classical solution, while the interesting short distance behaviour
is still maintained. This quantization procedure may provide a
general method for approximating the  quantum corrections to 
certain classical field configurations. 
\end{abstract}
\narrowtext

\section{Introduction}

Recently \cite{dzh1} several classical solutions to SU(3) Yang-Mills
theory were discussed, which possessed either spherical or cylindrical
symmetry. These solutions had gauge fields which tended toward
$\infty$ at large distances, leading to a type of confining behaviour
if one considered these solutions as background
fields in which some test particle moved. These increasing gauge
fields also led these solutions to the undesired property of
having infinite field energy. One way in which these
classical field configurations  might nevertheless have some 
physical importance is
if the quantization of these solutions reduced or eliminated
the bad long distance behaviour. While perturbative quantization
techniques work well for weakly interacting field theories such as
$QED$ (or $QCD$ in the high energy limit), they are not useful
when dealing with strongly interacting field theories. In Ref.
\cite{dzh2} we applied some ideas of Heisenberg's concerning the
quantization of strongly interacting, non-linear fields \cite{h1} to 
the cylindrical solution discussed in Ref. \cite{dzh1}, and
found that under certain assumptions the bad long distance 
behaviour of this solution was eliminated. Here we apply the
same procedure to the spherical symmetric solution and show
that again the bad long distance behaviour is eliminated.
In addition to the specific benefit that the Heisenberg
quantization method gives to the infinite energy solutions
discussed here and in Ref. \cite{dzh2}, it may provide some
general procedure for approximating the quantum corrections
to certain classical field configurations.

\section{Spherically symmetric ansatz}

We will briefly review the derivation and discuss some aspects of the
spherically symmetric solution.
The ansatz for the $SU(3)$ gauge field we take as in \cite{pagel}
\cite{palla} \cite{gal}:
\begin{mathletters}
\label{1}
\begin{eqnarray}
A _0 & = & \frac{2\varphi(r)}{{\bf i} r^2} \left( \lambda ^2 x - \lambda ^5 y 
      + \lambda ^7 z\right ) + \frac{1}{2}\lambda ^a
      \left( \lambda ^a _{ij} + \lambda ^a_{ji} \right ) 
      \frac{x^ix^j}{r^2} w(r),
\label{1a:1}\\
A^a_i & = & \left( \lambda ^a_{ij} - \lambda ^a_{ji} \right )
        \frac {x^j}{{\bf i} r^2} \left(f(r) - 1\right ) +
        \lambda ^a_{jk} \left (\epsilon _{ilj} x^k + 
	\epsilon _{ilk} x^j\right ) \frac{x^l}{r^3} v(r)
\label{1b:2},
\end{eqnarray}
\end{mathletters}
here $\lambda ^a$ are the Gell - Mann matrices; $a=1,2,\ldots ,8$
is a color index; the Latin indices $i,j,k,l=1,2,3$ are the space indices; 
${\bf i}^2=-1$; $r, \theta, \phi$ are the usual spherically coordinates. 
Substituting the ansatz of Eqs. (\ref{1}) (with $f = \phi =0$)
into the Yang - Mills equations 
\begin{equation}
\frac{1}{\sqrt{-g}}\partial _{\mu} \left (\sqrt {-g} {F^{a\mu}}_{\nu}
\right ) + f^{abc} {F^{b\mu}}_{\nu} A^c_{\mu} = 0,
\label{2}
\end{equation}
yields the following complex set of coupled, non-linear differential
equations \cite{pagel}
\begin{mathletters}
\label{2a}
\begin{eqnarray}
r^2f''& =& f^3 - f + 7fv^2 + 2vw\varphi - f\left (w^2 + \varphi ^2\right ), \\
r^2v''& = & v^3 - v + 7vf^2 + 2fw\varphi - v\left (w^2 + \varphi ^2\right ) \\
r^2w''& = & 6w\left (f^2 + v^2\right ) - 12fv\varphi, \\
r^2\varphi''& = & 2\varphi\left (f^2 + v^2\right ) - 4fvw.
\end{eqnarray}
\end{mathletters}
For the solution with increasing gauge fields we specialized
by taking $f= \varphi = 0$ (the case where $v = w =0$ is
similiar) gives the following set of non-linear coupled equations
\begin{mathletters}
\label{3}
\begin{eqnarray}
r^2 v''& = & v^3 - v - vw^2,
\label{3:1}\\
r^2 w''& = & 6w v^2.
\label{3:2}
\end{eqnarray}
\end{mathletters}
Near $r=0$ we took the series expansion form for $v$ and $w$ as
\begin{mathletters}
\label{4}
\begin{eqnarray}
v = 1 + v_2 \frac{r^2}{2 !} + ...,
\label{4:1} \\
w = w_3 \frac{r^3}{3 !} + ...
\label{4:2}
\end{eqnarray}
\end{mathletters}
where $v_2 , w_3$ were constants which determined the initial conditions
on $v$ and $w$ as in the last section.
In the asymptotic limit $r \rightarrow \infty$ the form of the solutions
to Eqs. (\ref{3}) approaches the form
\begin{mathletters}
\label{5}
\begin{eqnarray}
v & \approx & A \sin \left (x^{\alpha } + \phi _0\right ),
\label{5:1}\\
w & \approx & \pm\left [ \alpha  x^{ \alpha } + 
\frac{\alpha -1 }{4}\frac{\cos {\left (2x^{\alpha} + 2\phi _0 \right )}}
{x^{\alpha}}\right ],
\label{5:2}\\
3A^2 & = & \alpha(\alpha - 1).
\label{5:3}
\end{eqnarray}
\end{mathletters}
where $x=r/r_0$ is a dimensionless radius and $r_0, \phi _0$, and $A$ are 
constants. The second, strongly oscillating term in $w(r)$ is kept since
it contributes to the asymptotic behaviour of $w''$. We did not find
an analytical solution for the system of Eqs. (\ref{3}), but it is straight
forward to solve these equations numerically with any standard
differential equation package such as that available in
${\it Mathematica}$ \cite{math}. Fig. 1 shows a representative
solution to Eqs. (\ref{3}). The exponent of the power law increase
of $w$ (which is represented by $\alpha$ in the asymptotic
expressions) depended on the initial conditions, which were determined 
by the constants $v_2 , w_3$. Generally the exponent $\alpha$ would
decrease from a value in the range $2 -3$ to a value in the
range $1.2 - 1.8$ for a wide range of initial conditions. This behaviour
can be seen in the $Log (w)  - Log (x)$ plot in Fig. 2.
Although, these classical gauge fields
weakened slightly as r increased, they still diverged as
$r \rightarrow \infty$. Due to this feature of the ansatz function
$w$ the time part of the gauge field grew without bound as $r
\rightarrow \infty$, leading to both a classical type of confinement
(a test particle placed in the background field of this solution
would not be able to escape to $\infty$) and an undesired infinite
field energy for this solution. Various phenomenological studies of
quarkonia bound states use such increasing potentials to study the
spectrum of the bound state \cite{eich} although usually the potential
is taken to increase linearly. It should be mentioned that the
asymptotic form of the classical solution given in Eqs. (\ref{5})
are expected to be altered by the quantum corrections. The classical,
short distance behaviour, as given in Fig. 1, should be roughly
correct, since the pure gauge SU(3) theory that we are considering
is asymptotically free.
  
This ``bunker'' solution has ``magnetic'' and ``electric'' fields associated 
with it. Using the ansatz for $A_{\mu}$ from Eq. (\ref{1}) these 
``magnetic'' and ``electric'' fields have the following proportionalities 
\begin{mathletters}
\label{6}
\begin{eqnarray}
H^a _r & \propto & \frac{v^2-1}{r^2} , \; \; \; \; \; \; \; \;
H^a_{\varphi}  \propto  v' , \; \; \; \; \; \; \; \;
H^a_{\theta}  \propto  v' , \\
\label{6:1}
E^a_r & \propto & \frac{rw' - w}{r^2}, \; \; \; \; \; \; \; \;
E^a_{\varphi} \propto  \frac{vw}{r}, \; \; \; \; \; \; \; \; 
E^a_{\theta}  \propto  \frac{vw}{r},
\label{6:2}
\end{eqnarray}
\end{mathletters}
here for $E^a _r , H^a _{\theta}$, and $H^a_{\varphi}$ the color index 
$a=1,3,4,6,8$ and for $H^a _r, E^a _{\theta}$ and $E^a _{\varphi}$ 
$a=2,5,7$. The asymptotic behaviour of 
$H^a_{\varphi}, H^a_{\theta}$ and $E^a_{\varphi}, E^a_{\theta}$ 
is dominated by the strongly oscillating function $v(r)$. It
may be postulated that quantum corrections to this strongly 
ocsillating solution would tend to smooth it out so 
that it would not play a significant role in the large $r$
limit. From Eqs. (\ref{6}) and the asymptotic form of $v(r), w(r)$
the radial components of the ``magnetic'' and ``electric'' 
have the following asymptotic behaviour
\begin{equation}
\label{7}
H^a_r  \propto  \frac{1}{r^2}, \; \; \; \; \; \;
E^a_r  \propto  \frac{1}{r^{2-\alpha}}.
\end{equation}
where the strongly oscillating portion of $H^a _r$ is assumed not
to contribute in the limit of large $r$ due to smoothing by quantum
corrections. The radial ``electric'' field falls off slower than 
$1/r^2$ (since $\alpha > 1$) indicating the presence of a confining
potential. The $1/r^2$ fall off of $H^a _r$ indicates that this solution
carries a ``magnetic'' charge. This was also true for the simple solutions
discussed in Refs. \cite{pagel} \cite{palla}.
This leads to the result that if a test
is placed in the background field of the  bunker solution,
the composite system will have unusal spin properties ({\it i.e.} if the
test particle is a boson the system will behave as a fermion, and if the
test particle is a fermion the system will behave as a boson). This is
the spin from isospin mechanism \cite{rebbi}.

By examining the classical SU(3) field equations
of Eqs. (\ref{3}) we have found field configurations which led to a
classical confining behaviour, and which has some similarities with
certain phenomenological models used to study heavy quark
bound states. The most significant 
draw back of the present solutions is that it has infinite
field energy. The asymptotic form of the energy density goes as
\begin{equation}
{\cal E} \propto 4\frac{v'^2}{r^2} + \frac{2}{3}\left (
\frac{w'}{r} - \frac{w}{r^2}\right ) ^2 + 4\frac{v^2w^2}{r^4} +
\frac{2}{r^4} \left (v^2 - 1\right )^2 \approx \frac{2}{3}
\frac{\alpha ^2 (\alpha -1) (3 \alpha -1)}{x^{4-2\alpha}}
\label{8}
\end{equation}
Since we found $\alpha > 1$ this energy density will yield an
infinite field energy when integrated over all space. This can
be compared with the finite field energy monopole \cite{thooft}
and dyon solutions \cite{bps} \cite{julia}.

\section{Quantization of the ``bunker'' solution}
 
Although the classical confining behaviour of this ``bunker'' solution
may seem interesting due to its similiarity with certain phenomenological
potentials, the infinite field energy discussed at the end of the
previous section strongly argues against the physical importance
of this solution. One possible escape from this conclusion is if
quantum effects weakened or removed the bad long distance behaviour
of these solutions. However, strongly interacting,
non-linear theories are notoriously hard to quantize. In order to
take into account the quantum effects on the bunker solution
we employ a method used by Heisenberg \cite{h1} in attempts to
quantize the non-linear Dirac equation. By applying the dynamical
equation of motion (in Heisenberg's case the non-linear
Dirac equation) to an n-point Green's function, $G_n$, one would
arrive at an equation relating  $G_n$ to higher order
Green's functions ($G_{n+1}$ for example). Then applying the
dynamical equations of motion to the higher Green's functions
one would get equations relating these higher Green's functions
to even larger order Green's function. Continuing in this way
one arrived at an infinite set of differential equations relating
Green's functions of all orders. To handle this Heisenberg
employed the Tamm-Dankoff method whereby he only considered
Green's functions up to some order thus cutting off the infinite
set of equations. Here we will employ a similiar method to the
``bunker'' solution in terms of the ansatz functions $v$ and
$w$. Previously \cite{dzh2} we used this method on an infinite 
energy, string-like classical solution to the SU(3) equations.
For more details on the application of the Heisenberg method 
to such classical solutions we refer the reader to this article. 

In order to use
Heisenberg's quantization method on the nonlinear equations
we make the following assumptions :

1. The degrees of freedom relevant for studying the ``bunker''
solution (both classically and also quantum mechanically) are given
entirely by the two ansatz functions $w ,v$. No other
degrees of freedom arise through the quantization process.

2. From Eq. (\ref{5:2}) and Fig. 1 $w$ is a smoothly varying 
function for large
$x$, while $v$ is strongly ocsillating. Thus we take $w(x)$ to be
almost a classical degree of freedom while $v(x)$ is treated as a
fully quantum mechanical degree of freedom. One might think that
in this way only the behaviour of $v$ would change while $w$ stayed
the same. However since $w$ and $v$ are coupled via
the equations of motion we find that both functions
are modified.

To begin we replace the ansatz functions by operators ${\hat w} (x) ,
{\hat v} (x)$.
\begin{mathletters}
\begin{eqnarray}
\label{9}
x^2 {\hat v}'' &=& {\hat v}^3 - {\hat v} - {\hat v} {\hat w} ^2
\label{9:1} \\
x^2 {\hat w}'' &=& 6{\hat w} {\hat v}^2
\label{9:2}
\end{eqnarray}
\end{mathletters}
here the prime denotes a derivative with respect to $x$. Taking into
account assumption (2) we let ${\hat w} \rightarrow w$ become just
a classical function again, and replace ${\hat v}^2$ in Eq. ({\ref{9:2})
by its expectation value to arrive at
\begin{mathletters}
\begin{eqnarray}
\label{10}
x^2 {\hat v}'' &=& {\hat v}^3 - {\hat v} - {\hat v} w ^2 
\label{10:1} \\
x^2 w'' &=& 6 w \langle v^2 \rangle
\label{10:2}
\end{eqnarray}
\end{mathletters}
where the expectation value $\langle {\hat v}^2 \rangle$ is taken with
respect to fluctuations of $v$ around the classical solution
of Eq. (\ref{5}) or Fig. 1 ({\it i.e.} $\langle {\hat v}^2 \rangle = \int
{\cal D} v e^{i S /\hbar} v^2$ where $S$ is the action
and ${\cal D} v$ is a path integral measure over all possible 
configurations of $v$). If we took the expectation value of Eq. 
(\ref{10:1}) we would almost have a closed system of differential
equations relating $w$ and $\langle {\hat v} \rangle$. 
The $\langle {\hat v}^2 \rangle$
term from Eq. (\ref{10:2}) and the $\langle {\hat v}^3 \rangle$ term
from Eq. (\ref{10:1}) prevent the equations from being closed.
Applying the operation $x^2 \partial ^2 / \partial x ^2$ to the operator
${\hat v} ^2$ and using Eq. (\ref{10:1}) yields
\begin{equation}
\label{11}
x^2 ({\hat v ^2}) '' = 2 {\hat v} ^2 ({\hat v} ^2 -1 -w)
+ 2 x^2 ({\hat v}') ^2
\end{equation}
If we took the expectation of the above equation with respect
to fluctuations in the ansatz function operator ${\hat v^2}$,
and combined this with Eq. (\ref{10:2}) we would almost have
a closed system for determining $w$ and ${\hat v^2}$ except for
the $\langle ({\hat v}')^2 \rangle$ term which comes from the 
last term on the right hand side of Eq. (\ref{11}). Continuing
in this way one could obtain an infinite set of equations for
various powers of the ansatz function operator ({\it i.e.} 
${\hat v ^n}$). These higher order equations never close. 
To deal with this problem we follow
Heisenberg, and make some assumption that effectively cuts
off the system of equations at some finite order. By taking the
expectation of Eq. (\ref{11}) and further by assuming that
\begin{equation}
\label{12}
\langle ({\hat v'}) ^2 \rangle = {\langle {\hat v^2} \rangle
-v_0^2 \over x^2}
\end{equation}
we arrive at the closed system of equations from
Eqs. (\ref{10:2})  (\ref{11})
\begin{mathletters}
\begin{eqnarray}
\label{13}
x^2 \langle {\hat v}^2 \rangle '' &=& 2 \langle {\hat v}^2 \rangle ^2
-2 \langle {\hat v}^2 \rangle w - 2 v_0^2
\label{13:1} \\
x^2 w'' &=& 6 w \langle {\hat v}^2 \rangle
\label{13:2}
\end{eqnarray}
\end{mathletters}
By making the assumption of Eq. (\ref{12}) we
have simplified Eqs. (\ref{10:2}) (\ref{11}) to the
closed system given by Eqs. (\ref{13:1}) (\ref{13:2}). 
It is straightforward to show that in the limit $x \rightarrow
\infty$
\begin{mathletters}
\begin{eqnarray}
\label{14}
\langle {\hat v}^2 \rangle &=& v_0 ^2  + {a \over x ^{\alpha}}
\label{14:1} \\
w &=& {b \over x^{\alpha}}
\label{14:2}
\end{eqnarray}
\end{mathletters}
solves Eqs. (\ref{13:1}) (\ref{13:2}) 
provided that $v_0 ^2  = 1$, $b = -a$ and
$\alpha = 2 , -3$. In order for $\langle {\hat v}^2 \rangle$ and
$w$ to have acceptable behaviour at $x \rightarrow \infty$ we
take the $\alpha = 2$ solution. Substituting the above expressions
for $\langle {\hat v}^2 \rangle$ with $v_0 = +1$ and $w$ back into 
Eq. (\ref{10:1}), and assuming that $\langle {\hat v}^3 \rangle =
\langle {\hat v} \rangle \langle {\hat v}^2 \rangle$ gives the following 
equations for $\langle {\hat v} \rangle$ in the $x \rightarrow
\infty$ limit
\begin{equation}
\label{15}
x^2 \langle {\hat v} \rangle '' = \langle {\hat v} \rangle
(\langle {\hat v}^2 \rangle - 1) = \langle {\hat v} \rangle {a \over x^2}
\end{equation}
Eq. (\ref{15}) is solved by in the $x \rightarrow \infty$ limit
\begin{equation}
\label{15a}
\langle {\hat v} \rangle = \pm \left(1 + {a \over 6 x^2} \right)
\end{equation}
Eq. (\ref{15a}) together with Eqs. (\ref{14:1}) (\ref{14:2}) 
provide information on the behaviour of the ``classical'' ansatz
function, $w$, and the ``quantum'' ansatz function, $v$,
via $\langle {\hat v} \rangle$ and $\langle {\hat v}^2 \rangle$. The
main point of interest is that after applying the 
Heisenberg-like quantization procedure to the classical 
solution of Fig. 1, the infinite increase of
the ansatz function, $w$, has changed to an acceptable
asymptotic behaviour ({\it i.e.} one that leads to
a finite field energy). By replacing the $v^2$ and
$(v')^2$ terms in Eq. (\ref{8}) with $\langle {\hat v}^2 
\rangle$ and $(\langle {\hat v} \rangle ') ^2$ - from Eqs. 
(\ref{14:1}) and  (\ref{15a})  -
respectively, and also using $w$ from eq. (\ref{14:2}) we
find that the field energy density of the quantized ``bunker'' 
solution takes the form
\begin{equation}
\label{16}
{\cal E} \propto {a^2 \over x^8}
\end{equation}
in the limit in which quantum fluctuations become important
({\it i.e.} for non-Abelian theories which exhibit
asymptotic freedom this means in the low energy or
$x \rightarrow \infty$  range) the energy density goes from
the form given in Eq. (\ref{8}) to that given in Eq. (\ref{16}).
This can be seen to give a finite field energy. In the high
energy or short distance regime we assume that the 
fields approach the classical configuration of Figure 1
due to asymptotic freedom. This classical configuration
is well behaved at $x=0$, but would yield an infinite
field energy due to its divergence as $x \rightarrow \infty$.
In the long distance or low energy  limit 
the energy density should go over into the form 
given in Eq. (\ref{16}) which would then result in a finite
field energy for this configuration, since the integral of
Eq. (\ref{16}) over  the large $x$ region, where it is
valid, would give a finite field energy.

Finally by using $w$, $\langle {\hat v} \rangle$, and $\langle
{\hat v}^2 \rangle$ in the expressions for ${\bf E}$ and ${\bf H}$
given in Eq. (\ref{6}) we find that the radial
fields ($E_r , H_r$) go like $a / r^4$ while the
angular fields ($E_{\theta , \phi} , H_{\theta , \phi}$)
go like $a / r^3$. Thus the quantization procedure
outlined above modifies the undesirable long distance
behaviour of the ``electric'' and ``magnetic'' fields
as well as the energy density.

\section{Discussion}

In this paper we reviewed a certain classical field
configuration for an SU(3) gauge theory.
Near the origin the field configurations were finite, but
as $r \rightarrow \infty$ the fields diverged (see Figs. 1 ,2).
This increasing field strength led to a classical type
of confinement in that a test particle placed in the background 
field of this solution would not be able to escape to
$\infty$. Unfortunately, this diverging of the field
as $r \rightarrow \infty$  also led to this configuration having 
an infinite field energy. Previously it was suggested that
quantum effects might soften or eliminate this bad long
distance behaviour. By applying a method similiar to that
Heisenberg used in quantizing the non-linear Dirac
equation we find that the long distance behaviour is
changed so as to give finite field energy. At short distances
the fields should approach the classical configuration of
Fig. 1 from the asymptotic freedom of the SU(3) gauge 
theory. This classical solution has the good features of not being
divergent at $r=0$ and in some limited region around
$r=0$ the fields increase in a way similiar to that
found in some phenomenological models of confinement.
At long distances the fields should approach the 
configuration given by Eqs. (\ref{14:1}) (\ref{14:2})
(\ref{15a}) where the
quantum effects have eliminated the divergence of
the fields and field energy density as $r \rightarrow
\infty$.

\newpage
\centerline{{\bf List of figure captions}}

Fig.1. The $w(x)$ confining function, and the $v(x)$ oscillating function 
of the $SU(3)$ bunker solution. The initial conditions for this 
particular solution were $v_2 = 0.1$, $w_3 = 2.0$, and $x_i = 0.001$.

\vspace{1.0in}

Fig.2. A plot of $Log(w) - Log(x)$ of the solution from Fig. 2 showing 
the different power law behaviour in the small $x$ and large $x$ regions.

\end{document}